\documentclass[prd,twocolumn,showpacs, preprintnumbers,amsmath,amssymb,floatfix]{revtex4}      
\usepackage{epsfig}%
\usepackage{latexsym}
\usepackage{bm}
\usepackage{amssymb}
\usepackage{graphicx}
\usepackage{rotating}
\usepackage{longtable}
\usepackage{revsymb}
\usepackage{dcolumn}
\begin{document}
\setlength{\topmargin}{-0.25in}
\preprint{U. of IOWA Preprint}

\title{Universality in nontrivial continuum limits: a model calculation}

\author{Y. Meurice}
\email[]{yannick-meurice@uiowa.edu}
\affiliation{Department of Physics and Astronomy\\ The University of Iowa\\
Iowa City, Iowa 52242 \\ USA
}

\author{M. B. Oktay}
\email[]{oktay@uiuc.edu}
\affiliation{Department of Physics\\ The University of Illinois\\
1110 West Green Street \\ Urbana, IL 61801\\ USA
}


\date{\today}

\begin{abstract}
We study numerically the continuum limit corresponding to the non-trivial fixed point of 
Dyson's hierarchical model.
We discuss the possibility of using 
the critical amplitudes as input parameters. 
We determine numerically the leading and subleading critical amplitudes of the zero-momentum 
connected $2l$-point functions in the symmetric phase up to the 20-point function for 
randomly chosen local measures. 
Using these amplitudes, we construct quantities which are expected to be universal 
in the limit where very small log-periodic corrections are neglected: the $U^{(2l)\star}$
(proportional to the connected $2l$-point functions) and the $r_{2l}$ (proportional to one-particle irreducible(1PI)). 
We show that these quantities are independent of the 
the local measure with at least 5 significant digits. We provide clear evidence for the asymptotic behavior  $U^{(2l)\star}\propto (2l)!$ and reasonable evidence for 
$r_{2l}\propto (2l)!$. These results signal a finite radius of convergence for the 
generating functions. We provide numerical evidence for a linear growth for 
universal ratios of subleading amplitudes. We compare our $r_{2l}$ with existing 
estimates for other models.
\end{abstract}


\maketitle

\section{Introduction}
Predictive theories are highly regarded because they offer many ways to be falsified. On the 
other hand, theories where every new observation results into the determination 
of a new parameter are uninteresting unless this process ends up with predictions. 
The condition of perturbative
renormalizability allows us to get rid of UV regulators without a proliferation of adjustable parameters. It played a central role in the establishment of the 
standard model of electroweak and strong interactions.
This model predicts many cross sections and decay rates 
\cite{pdg02} in terms of a very restricted set of input parameters and is considered 
as one of the most important accomplishment of the 20-th century. 

In this article, we address the question of predictiveness for the infinite cutoff limit
associated 
with a nontrivial fixed point. This concept was developed with a 3 dimensional example by 
Wilson \cite{wilson72}. In this case, we are free to use 
a much larger set of bare theories, including non-renormalizable interactions but we need to 
fine-tune one of the parameters in order 
to reach the critical hypersurface, or stable manifold, which separates the ordered and disordered phases.
The renormalization group (RG) flows go along the stable manifold until they get close to the 
non-trivial fixed point, and then end up along the unstable manifold. 
As the fined-tuned parameter gets close to its critical value, we can parametrize the 
zero momentum 
$q$-point functions in terms of power singularities multiplied by critical amplitudes.
A few years ago \cite{gam3rapid}, we have suggested to use (some of) the critical
amplitudes as input parameters. 
However, it requires that we know how to 
pick an independent set of these amplitudes. 

Before the nineties \cite{privman91} this question has not generated much interest. However, 
in the last decade, universal ratios of amplitudes associated with 
one-particle irreducible $2l$-point functions at zero momentum
have been calculated with various methods \cite{Tetradis94,Tsypin94,Guida97,Morris97,Pelissetto98b,Campostrini99,Campostrini00} 
(see Ref. \cite{Pelissetto00} for a review and a more extended list of references). 
These ratios are denoted $r_{2l}$ as in Ref. \cite{Campostrini99} where Table VIII 
summarizes the values of $r_6, \ r_8$ and $r_{10}$ obtained in the literature. 
We are only aware of one calculation \cite{Morris97} of $r_{12}$ and $r_{14}$ with 
large error bars. As we will see it is difficult to extrapolate the asymptotic behavior of 
the $r_{2l}$  from this data.

In this article, we calculate universal quantities associated with the $2l$-point functions
in a model where very accurate calculations of these quantities
are possible up to the 20-point function.
We use Dyson's hierarchical model \cite{dyson69,baker72} which is very 
close to the 3 dimensional model used in Ref. \cite{wilson72}. For definiteness, this model and our method of calculation 
are reviewed in section \ref{sec:model}. 
In section \ref{sec:amp}, we discuss the 
numerical calculation of the critical amplitudes for the 
zero-momentum 
$2l$-point functions. The definition of the renormalized couplings in terms of these amplitudes 
is discussed in section \ref{sec:npr}, where we also explain why we expect some dimensionless 
quantities made out of these couplings to be approximately universal \cite{smalld03}.
In section \ref{sec:lead}, we verify numerically that our expectations are 
correct and that in the infinite 
cutoff limit, dimensionless renormalized couplings \cite{parisi88} are in good approximation 
universal. 
The results  presented here extend up to the 20-point function and allow us to study 
the asymptotic behavior. 
We found good evidence for a factorial growth comparable 
to what is found in Refs. \cite{goldberg90,cornwall90,zakharov91,voloshin92} 
for other models studied in the context of 
multiparticle production.

In section \ref{sec:sublead}, we show that
the ratios of subleading 
amplitudes are in good numerical approximation universal 
for the model considered here, as expected in general\cite{wegner76,aharony80,chang80}.
In section \ref{sec:effpot}, we calculate the $r_{2l}$, show that they are universal and of the same order 
of magnitude as those calculated 
in the literature for other models. Their asymptotic behavior is compatible with a factorial growth as 
suggested by ``Griffiths analyticity"\cite{Campostrini99}.
In the conclusions, we discuss the interpretation of these results and the 
applicability of the method to more realistic situations.

\section{The model}
\label{sec:model}

Dyson's hierarchical model \cite{dyson69,baker72} 
has a special kinetic term that keeps its original form after a RG 
transformation. There is no wave-function 
renormalization in this model and $\eta$=0. The RG transformation can be 
summarized by a simple integral formula that describes the change in the local measure 
after $n$ RG steps $W_n(\phi)$, into the one after one more step:
\begin{eqnarray}
W_{n+1}(\phi) =
C_{n+1}\exp((\beta/2)\phi^2)\nonumber \\
\times \int
d\phi^{'}W_n({(\phi-\phi^{'})c^{1/2}})W_n({(\phi+\phi^{'})c^{1/2}})\ ,
\end{eqnarray}
where $C_{n+1}$ is a normalization factor which can be fixed at our
convenience. The model has one free parameter $c=2^{1-2/D}$ which can be 
adjusted in order to match the scaling of a free massless field in $D$ dimensions.
In the following we use the value $c=2^{1/3}$ corresponding to $D=3$ exclusively.
For this value of $c$, 
the non-trivial fixed point is known with great 
accuracy \cite{koch95}. 
A list of accurate values of the eigenvalues of the linear RG 
transformation about this fixed point is given in Ref. \cite{gam3}.
In the following, these eigenvalues are denoted $\lambda_i$ with $\lambda_1\simeq 1.427 $
the largest and only relevant eigenvalue.
We restrict our investigation to the symmetric phase 
of the model and use the 
accurate methods of calculations \cite{finite} available in this phase.

This integral formula is very similar to 
the approximate recursion 
formula \cite{wilson71b} used in Ref. \cite{wilson72}. 
The main difference is that we integrate 2 field variables (keeping 
their sum constant) in one RG step instead of $2^D$ field variables.
As a result, the change in the linear scale of the blocks is $2^{1/D}$ (instead of 2).
This difference has no effect from a qualitative point of view.
For a quantitative comparison of the two cases see Ref. \cite{fam}.
A more explicit description of Dyson's hierarchical model and a more complete list of references can be found 
in Ref. \cite{gam3}. 

The RG transformation defines a flow in the space of local measures 
$W(\phi)=\exp (-V(\phi))$. We require parity invariance and $V\rightarrow +\infty$ 
when $|\phi|\rightarrow +\infty$ at a quadratic rate or faster.
In the following calculations, 
the bare parameters will appear in a local measure
of the Landau-Ginzburg (LG) form:
\begin{equation}
W_0(\phi)\propto exp^{-({1/2}m^2 \phi^2+ g\phi^{2p})}\ .
\label{eq:lg}
\end{equation}
We have considered the 
(randomly chosen) possibilities given explicitly in Table \ref{table:measures} provided 
in the Appendix. One of the choices is $p=4$ and corresponds to non-renormalizable 
interactions in perturbation theory.
In addition, we considered the Ising measure $W(\phi)=\delta(\phi^2-1)$. For 
each local measure, we have varied the inverse temperature $\beta$ in front of the 
kinetic term in order to reach a critical value $\beta_c$ where a bifurcation 
is observed (see \cite{gam3rapid} for a more complete description). 

We have then studied numerically the flows of the local measures for values of 
$\beta $ slightly smaller than their respective $\beta_c$. 
After $n$ RG steps, the 
Fourier transform of the local measure
\begin{equation}
R_n(k)=1+a_{n,1}k^2+a_{n,2}k^4+\dots  \ ,
\end{equation}
contains the information concerning the zero-momentum $q$-point functions.
The connected parts \cite{smalld03} are defined by 
\begin{equation}
\ln (R_n(k))=a^c_{n,1}k^2+a^c_{n,2}k^4+\dots  \ ,
\end{equation}
with
\begin{equation}
a^c_{n,l}=(-1)^l \frac{1}{2l!}(\frac{c}{4})^{ln}\langle (\sum_{2^n sites} \phi_x ) ^{2l}\rangle ^c \ .
\label{eq:fulla}
\end{equation}
We can then calculate 
\begin{equation}
\chi^{(2l)}_n\equiv\frac{\langle \left(\sum_{2^n {\text sites}} \phi_x  \right)   ^{2l}\rangle ^c}{2^n}
\end{equation}
in terms of the $a_{n,l}$. For instance,
\begin{equation}
\chi^{(4)}_{n} = 12\;(-a_{n,1}^2+2 a_{n,2})(8/c^2)^n \ ,
\end{equation}
and 
\begin{equation}
\chi^{(6)}_{n} = 240\;(a_{n,1}^3-3\;a_{n,1}\;a_{n,2}+3\;a_{n,3})(32/c^3)^n \ .
\end{equation}
For $\beta<\beta_c$ the $\chi_n^{(2l)}$ have a well-defined limit when $n$ becomes large 
that we call $\chi^{(2l)}$ and that we now proceed to parametrize.

\section{Numerical determination of the critical amplitude}
\label{sec:amp}

For values of 
$\beta $ slightly smaller than $\beta_c$, we have 
\begin{eqnarray}
\chi^{(2l)} \simeq &\ & ( \beta_c-\beta )^{-\gamma_{2l}} \left[ A_0^{(2l)}+A_1^{(2l)}
\left(\beta_c-\beta \right)^\Delta \right.  \nonumber\\
&+& \left. A^{(2l)}_{per.}\cos \left(\omega \ln (\beta_c-\beta)+\phi ^{(2l)} \right)+ \dots  \right] ,
\end{eqnarray}
with known parameters $\gamma_{2l}$, $\Delta$ (not to be confused with the gap 
exponent) and $\omega$.
We use the hyperscaling \cite{finite,hyper} values 
\begin{equation}
\gamma_{2l}=\gamma (5l-3)/2 \ ,
\label{eq:hyper}
\end{equation}
with 
\cite{gam3}
\begin{equation}
\gamma \simeq 1.299140730159
\end{equation}
and 
\begin{equation}
\Delta \simeq 0.42595\ .
\end{equation}
The possibility of log-periodic terms 
were first discussed in Refs. \cite{wilson72,
niemeijer76}. They were 
identified in the high-temperature expansion \cite{osc1,osc2} of the model considered 
here. The frequency is 
\begin{equation}
\omega =\frac{2\pi}{\ln \lambda	_1}\ .
\end{equation}
The amplitudes $A^{(2l)}_{per.}$ 
are however quite small, typically, they affect the 16-th significant 
digit of the susceptibility and it takes a special effort to resolve them 
numerically.

As a starting point, we will calculate the subleading amplitude $A_1^{(2l)}$. Using
the parametrization $\beta= \beta_c-10^{-x}$ and a small $\delta x$,
we obtain (neglecting the log-periodic corrections) 
\begin{equation}
A_1^{(2l)}\simeq\frac {[\chi^{(2l)}(x+\delta x)10^{-(x+\delta x)\gamma_{2l}}-\chi^{(2l)}(x)10^{-x\gamma_{2l}}]}
{10^{-x\Delta}[10^{-\Delta\delta x}-1]}
\vspace{12pt}
\end{equation}
As $x$ becomes large enough, we observe a region where $A_1^{(2l)}$ stabilizes.  
Once $A_1^{(2l)}$ are obtained, it is easy to find $A_0^{(2l)}$
from
\begin{equation}
A_0^{(2l)}=\chi^{(2l)}\;(\beta_c-\beta)^{\gamma_{2l}}-A_1^{(2l)}(\beta_c-\beta)^\Delta.
\end{equation}
More details regarding the numerical 
methods can be found in Ref. \cite{oktayphd}. 

\section{Non-perturbative renormalization}
\label{sec:npr}
In this section, we give a non-perturbative definition of the renormalized couplings.
In the next section, we discuss the relation between these couplings and the critical 
amplitudes.
We follow the general procedure outlined by Wilson in Ref. \cite{wilson72}.
We consider a sequence $L=1,2\dots$ of models with 
$\beta=(\beta _c  -\lambda_1 ^{-L} u)$
where 
$u$ is positive but not too large.   
We introduce the increasing sequence of UV cutoffs 
\begin{equation}
\Lambda_L=2^{\frac{L}{D}}\Lambda_R \ , 
\end{equation}
with $\Lambda_R$ an arbitrary scale of reference. 
We follow the notations of Ref. \cite{wilson72} and $L$ which is 
proportional to the logarithm of the UV cutoff 
should not be confused with 
the linear size of the system.
We define the renormalized mass
\begin{equation}
m_R^2 =\frac{\Lambda_L^2}
{\chi^{(2)}(\beta _c  -\lambda_1 ^{-L} u)}\ .
\end{equation}
Given that
\begin{equation}
\lambda_1^\gamma ={2^{\frac{2}{D}}},
\end{equation}
the dependence on the UV cutoff disappear at leading order and one obtains
\begin{equation}
m_R^2=\frac{\Lambda_R^2 u^\gamma}{A^{(2)}_0+A^{(2)}_1u^{\Delta}(\frac{\Lambda_R}
{\Lambda_L})^{\frac{2\Delta}{\gamma}}+LPC+\dots}
\end{equation}
with the log-periodic corrections 
\begin{equation}
LPC=A^{(2)}_{per.}\cos \left(\omega \left(\ln u+\frac{2}{\gamma}\ln \left(\frac{\Lambda_R}{\Lambda_L}\right)\right)+\phi^{(2)}\right) \ .
\end{equation}
In the infinite cut-off limit $(L\rightarrow\infty)$, the subleading corrections 
disappear. On the other hand, the LPC do not and we are in presence of a limit 
cycle with a cutoff dependence quite similar to Refs. \cite{wilson03,braaten03}.
Strictly speaking the infinite cutoff limit does not exists, however, for practical 
purpose, the effects of the oscillations are so small that it introduces uncertainties
that are smaller than the accuracy with which we establish the universality.
Consequently, these oscillations will be ignored in the following.

We can now define $U^{(q)}$, the dimensionless coupling constants \cite{parisi88} associated with the $q$-point
function as
\begin{equation}
U^{(q)}\propto \chi^{(q)}(\beta _c  -\lambda_1 ^{-L} u)\;m_R^{q(1+D/2)-D}\ .
\end{equation}
The constant of proportionality is fixed in such way that 
if the conjecture \cite{glimm87} that $ (-1)^{l+1}\chi^{(2l)}>0 $ is 
correct, then $U^{(2l)}>0$.
We also introduce a power of $\beta$ such that if we reabsorb $\beta$ in the 
field definition, we get comparable quantities for 
measures with different $\beta_c$.
In summary, for $D=3$
\begin{equation}
U^{(2l)}\equiv (-1)^{l+1}\chi^{(2l)}\;(\chi^{(2)})^{(3-5l)/2}\;\beta^{3(1-l)/2} \ .
\label{eq:lamdastar}
\end{equation}
In this definition, it is understood that $\chi^{(2l)}$ and $\chi^{(2)}$ are evaluated 
at the same $\beta$. 

In section X of Ref. \cite{smalld03}, it is shown that when $\beta \rightarrow \beta_c$ (and the RG flows 
end up on the unstable manifold), the $U^{(2l)}$ are universal in the approximation where the log-periodic 
oscillations are neglected. Conversely, if we find that in this limit, the $U^{(2l)}$ are 
approximately universal, it indicates that the log-periodic 
oscillations are small.

\section{Universal couplings}
\label{sec:lead}

We now consider the infinite cutoff limit $U^{(2l)\star}$ of $ U^{(2l)}$.
In this limit, the subleading corrections vanish.
Due to the hyperscaling relation Eq. (\ref{eq:hyper}), the leading singularities cancel
and we obtain
\begin{equation}
U^{(2l)\star}=(-1)^{l+1}A_0^{(2l)}\;(A_0^{(2)})^{(3-5l)/2}\;\beta_c^{3(1-l)/2}.
\end{equation}
\noindent
Using the previously calculated amplitudes
we find that up to $l=10$, the $U^{(2l)\star}$ are 
all positive and in good approximation universal. The values for particular measures are given in 
the Appendix (Table
\ref{table:ustar}).
The approximately universal values are displayed in Table \ref{table:uq} with uncertainties 
of order 1 in the last digit.

\begin{figure}[ht]
\includegraphics[width=3.2in,angle=0]{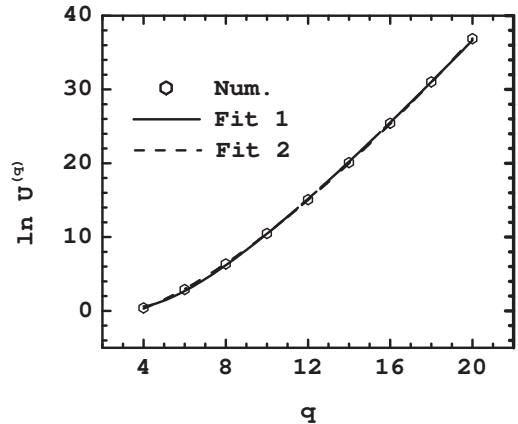}
\caption{$\ln (U^{(q)\star})$ and the two fits described in the text.
\label{fig:urat}}
\end{figure}

We have fitted $\ln U^{(q)\star}$ with a constant plus a linear term and a third term which is either $\ln q$ or $\ln (q!)$.
Fig. \ref{fig:urat} shows two fits of these 9 values.
The first fit (Fit 1 in Fig. \ref{fig:urat}) is 
\begin{equation}
U^{(q)\star}\simeq 21.5\exp(3.436q)q^{-11.7}\ ,
\end{equation}
and the second fit (Fit 2 in  Fig. \ref{fig:urat})
\begin{equation}
U^{(q)\star}\simeq 0.756 (q !)^{1.29}\exp(-0.88 q)
\end{equation}
\begin{table}
\caption{\label{table:uq}Universal values of $U^{(2l)\star}$ .}
\begin{ruledtabular}
\begin{tabular}{cc}
\hline
2l & $U^{(2l)\star}$ \\
\hline
4 & 1.505871 \\ 6 & 18.10722 \\ 8 & 579.970 \\ 10 & 35653.8 \\
 12 & ${\displaystyle 3.57769\, {10}^6 }$ \\ 14 & ${\displaystyle 5.31763\,{10}^8}$ \\ 
 16 & ${\displaystyle1.09720\, {10}^{11}}$ \\ 
 18 & ${\displaystyle 3.00025\,{10}^{13} }$\\ 20 & ${\displaystyle 1.04998\,{10}^{16}}$ \\  
\end{tabular}
\end{ruledtabular}
\end{table}  

\begin{figure}
\includegraphics[width=3.2in,angle=0]{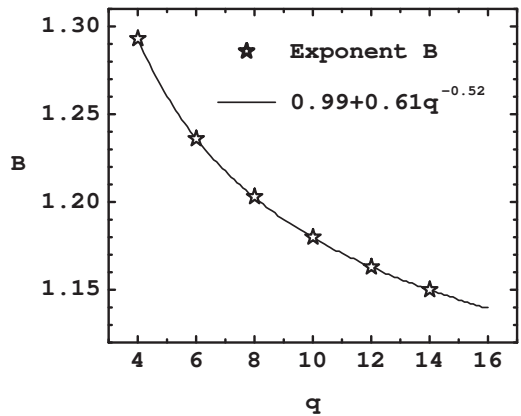}%
\caption{Value of the fitted exponent of the factorial for different sets of data.
The abscissa is the smallest value of $q$ used in the data set. \label{fig:bfit}}
\end{figure}

The two fits can barely be distinguished in Fig. \ref{fig:urat}. 
If as in Fit 1 we use a general fit 
of the form 
\begin{equation}
U^{(q)\star}\simeq A(q !)^{B}\exp(C q)\ ,
\end{equation}
then the values of the parameters change if we exclude the points with 
low values of $q$. For instance if we  exclude the first five points, 
$B\simeq 1.15$ instead of 1.29 if we use all the data.
The intermediate values are shown on Fig. \ref{fig:bfit}. Using a nonlinear 
fit to determine how $B$ depends on this choices, 
we found that $B \simeq 0.99+0.61q^{-0.52}$ where $q$ is the smallest index in the 
dataset used to obtain $B$. 
We conclude that the asymptotic value is very close to 1 and that the the leading 
growth is 
\begin{equation}
U^{(q)\star}\approx q !   \ ,
\end{equation}
This result is similar 
to what is found in Ref.  \cite{goldberg90,cornwall90,zakharov91,voloshin92}
for other models studied in the context of 
multiparticle production. Note that the generating function of the connected 
$2l$-points function has a $1/(2l)!$ factor at order $2l$ (see Eq. (\ref{eq:fulla})) 
which indicates that the expansion of the generating function of the connected functions in powers of an 
external field has a finite radius of convergence (see section 
\ref{sec:effpot} for more discussion).
\section{Correction to Scaling Amplitudes}
\label{sec:sublead}
\begin{figure}[!t]
\vskip10pt
\includegraphics[width=3.2in,angle=0]{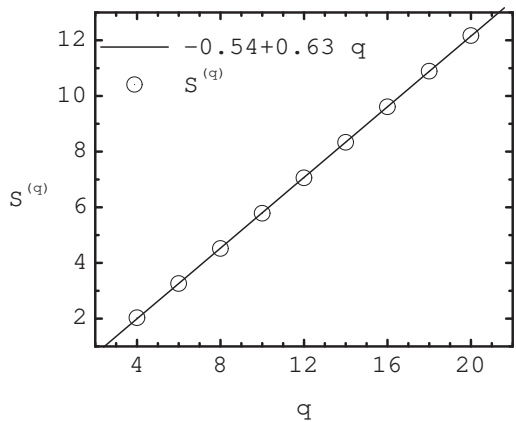}%
\caption{Ratios of subleading amplitudes $S^{(q)}$ \label{fig:subl}}
\end{figure}

Ratios of subleading amplitudes are also expected to 
be universal \cite{wegner76,aharony80,chang80}.
To express these quantities we first define
the relative strength of the corrections
$a^{(q)}\equiv A_1^{(q)}/A_0^{(q)}$. From these, we define the 
ratios 
\[ S^{(q)\star}=\frac{a^{(q)}}{a^{(2)}} \ . \]
For the measures considered here, we found the approximately universal values 
shown in Table \ref{table:sq} with uncertainties of order 1 in the last digit. 
Particular values are given in the Appendix.
\begin{table}
\caption{\label{table:sq}Universal values of $S^{(2l)\star}$ .}
\begin{ruledtabular}
\begin{tabular}{cc}
\hline
2l & $S^{(2l)\star}$ \\
\hline
4 & 2.03 \\ 6 & 3.26 \\ 8 & 4.52 \\ 10 & 5.8 \\ 12 & 
   7.1 \\ 14 & 8.3 \\ 16 & 9.6\\ 18 & 11\\ 20 & 
   12 \\ 
\end{tabular}
\end{ruledtabular}
\end{table}  
These values can be fitted well with a linear function as shown in Fig. \ref{fig:subl}
\section{The effective potential}
\label{sec:effpot}
In this section, we compare our results with related results obtained for models in the 
same class of universality 
as the 3 dimensional Ising model with nearest neighbor interactions. In order to 
avoid repetitions, we follow sections V, VII and Appendix C of Campostrino, Pelissetto, Rossi and Vicari \cite{Campostrini99} (CPRV for short). We follow 
exactly their notations except for the fact that we keep using superscripts 
in $\chi^{(2q)}$ to denote the connected $2q$-points function (denoted  $\chi_{2q}$ in CPRV). It should also be mentioned that in the case of Dyson's model, there is no wave function 
renormalization and consequently some of the rescalings introduced in CPRV are not 
necessary in the present case.

In CPRV, a universal function 
\begin{equation}
A(z)=z^2/2+z^4/24+\sum_{l\geq 3}\frac{r_{2l}}{(2l)!}z^{2l}\ ,
\label{eq:aofz}	
\end{equation}
obtained from the effective potential by suitable rescalings of the 
effective potential and its argument (the magnetization) is defined (Eq. (5.7)). 
The universal coefficients $r_{2l}$ 
are ratios of amplitudes associated with the 1PI. They can be expressed in terms of 
the connected Green's functions by expanding the magnetization as a power series of the magnetic field 
in the equation defining the Legendre transform. Explicit formulas for $r_6$, $r_8$ and $r_{10}$ 
are given in Eqs. (5.23-25) of CPRV. These quantities 
can be trivially reexpressed
section \ref{sec:lead}. For instance, 
\begin{equation}
r_6=10-\frac{U^{(6)\star}}{(U^{(4)\star})^2}\ .
\end{equation}
In CPRV, ``Griffiths' s analyticity''is invoked  to justify 
that $A(z)$ in Eq. (\ref{eq:aofz}) has a finite radius of convergence. 
This requires, for large $l$, a growth not faster than
\begin{equation} 
r_{2l}\propto (C_1)^l(2l)! \ .
\label{eq:rrate}
\end{equation}

The coefficients of $r_{2l}$ grow rapidly as can be observed in 
\begin{eqnarray}
r_{12}&=&1401400 - 560560 \frac{\chi^{(2)}\chi^{(6)}}{{\chi^{(4)}}^2}
+36036\frac{{\chi^{(2)}}^2{\chi^{(6)}}^2}{{\chi^{(4)}}^4}
\nonumber \\
   &+&
 17160\,\frac{{{{\chi}^{(2)}}}^2\,{{\chi}^{(8)}}}{{{{\chi}^{(4)}}}^3} - 
 792\, \frac{{{{\chi}^{(2)}}}^3\,{{\chi}^{(6)}}\,{{\chi}^{(8)}}}
   {{{{\chi}^{(4)}}}^5} \nonumber \\ &-& 220\,\frac{{{{\chi}^{(2)}}}^3\,{{\chi}^{({10})}}}
   {{{{\chi}^{(4)}}}^4}  + \frac{{{{\chi}^{(2)}}}^4\,{{\chi}^{({12})}}}
   {{{{\chi}^{(4)}}}^5}\ ,
\end{eqnarray}
and
\begin{eqnarray}
 r_{14} &=&  190590400 - 95295200\,\frac{{{\chi}^{(2)}}\,{{\chi}^{(6)}}}
   {{{{\chi}^{(4)}}}^2} \nonumber \\ &+& 10090080\,\frac{{{{\chi}^{(2)}}}^2\,{{{\chi}^{(6)}}}^2}
   {{{{\chi}^{(4)}}}^4}  -126126\,\frac{{{{\chi}^{(2)}}}^3\,{{{\chi}^{(6)}}}^3}
   {{{{\chi}^{(4)}}}^6} \nonumber \\ &+& 3203200\,\frac{{{{\chi}^{(2)}}}^2\,{{\chi}^{(8)}}}
   {{{{\chi}^{(4)}}}^3} - 360360\,\frac{{{{\chi}^{(2)}}}^3\,{{\chi}^{(6)}}\,
     {{\chi}^{(8)}}}{{{{\chi}^{(4)}}}^5} \nonumber \\ &+& 
 1716\, \frac{{{{\chi}^{(2)}}}^4\,{{{\chi}^{(8)}}}^2}{{{{\chi}^{(4)}}}^6} - 
  50050\,\frac{{{{\chi}^{(2)}}}^3\,{{\chi}^{({10})}}}{{{{\chi}^{(4)}}}^4}\nonumber \\ & +& 
  2002\,\frac{{{{\chi}^{(2)}}}^4\,{{\chi}^{(6)}}\,{{\chi}^{({10})}}}
   {{{{\chi}^{(4)}}}^6} + 364\,\frac{{{{\chi}^{(2)}}}^4\,{{\chi}^{({12})}}}
   {{{{\chi}^{(4)}}}^5} \nonumber \\ &-& \frac{{{{\chi}^{(2)}}}^5\,{{\chi}^{({14})}}}
   {{\chi{(4)}}^6}\ .
   \end{eqnarray}
The first (constant) and last 
(proportional to $\chi^{(2l)}$) terms of the $r_{2l}$ 
are expected to grow like $(2l)!$. The first term of the $r_{2l}$ (10, 280, 15400, 1401400, etc...) 
denoted $r^{(0)}_{2l}$, can be obtained from the recursion 
\begin{equation}
	r_{2l}^{(0)}=\sum_{m=1}^{[2l/3]}r_{2(l-m)}^{(0)}\frac{(-1)^{m+1}}{(2l-3m)!m!6^m}\ 
\end{equation}
with the initial conditions $r_2^{(0)}=r_{4}^{(0)}=1$. 
A detailed analysis shows that this leads to a $(2l)!$ growth (with power corrections). 
On the other hand,  we also expect $\chi^{(2l)}\approx (2l)!$ in view of the numerical results of 
section \ref{sec:lead}. Obviously, if similar rates are found for the intermediate 
terms, then the bound of Eq. (\ref{eq:rrate}) applies.

The numerical values of $r_6, \dots r_{20}$ for the four measures are given in 
the Table \ref{table:r2l} in the appendix. The results show universality with the 
same kind of accuracy as in Section \ref{sec:lead}. The universal values are 
summarized in Table \ref{table:rl}. 
\begin{table}
\caption{Universal values of $r_{2l}$ calculated numerically and 
predicted using the CPRV method for the value of $\rho$ given in the last column. \label{table:rl}}
\begin{ruledtabular}
\begin{tabular}{cccc}
\hline
2l & $r_{2l}$ & $r_{2l}$ predicted&$\rho$ \\
\hline
 6 & 2.0149752 & 1.81& 1.8946\\ 
 8 & 2.679529 & 2.47 & 1.8946\\ 
 10 & -9.60118 & -10.1&1.9218\\ 
 12 & 10.7681 &8.93&1.9460\\
 14 & 763.062& 753&1.9685\\
 16&-18380.8& $-1.76\times 10^4$& 2.3197\\
 18&$1.5553\times 10^6$& $1.62\times 10^6$ & 2.1889\\
 20&$1.0374 \times 10^7$& $1.04 \times 10^7$& 2.1181
\end{tabular}
\end{ruledtabular}
\end{table}  
It should be noted that the 
numerical values of the $r_{2l}$ are orders of magnitudes smaller than some of 
the individual terms. For instance, for $r_{14}$, the sum of all the terms $(753)$ is 5 orders of magnitudes smaller than the constant term ($1.9 \times 10^8$). 
This requires minute cancellations, and since from the discussion above, some of the terms grow as $(2l)!$, this is a good indication that the bound of Eq. (\ref{eq:rrate}) should be satisfied.
  
We have checked the consistency of our results by using the method proposed by 
CPRV to predict $r_{2(l+1)}$, using $r_2, \dots r_{2l}$ as input.
As one can see from Table \ref{table:rl}, the two results are in reasonable agreement (relative errors of the order of 10 percent).
One can also calculate 
$r_{2(l+2)}$ etc.. from the same input, but the agreement is not 
as good. It should 
noted that Eq. (7.21) in CPRV, for the intermediate parameter $\rho$, admits in general more than one positive root. The numbers given  here have been obtained 
by using the smallest positive root also given in Table \ref{table:rl}.  
We also found that the equations leading to the prediction of $r_6$ and $r_8$ 
were identical. A detailed analysis shows that it is due to the fact
that the magnetization exponent 
(usually denoted $\beta$) is $\gamma/4$ \cite{hyper} for Dyson's model .

\begin{figure}
\vskip60pt
\includegraphics[width=3.2in,angle=0]{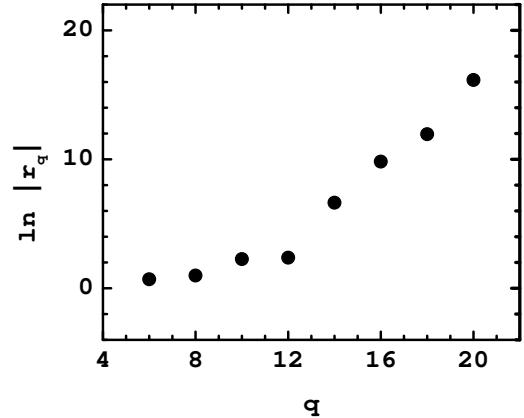}%
\vskip-60pt
\caption{Values of $\ln |r_q|$ versus $q$ \label{fig:lnrl}}
\end{figure}

The logarithm of $|r_{2l}|$ is displayed in  Fig. \ref{fig:lnrl}. The growth looks roughly similar to the 
growth of the $U^{(2l)\star}$ shown in Fig. \ref{fig:urat}, however 
the behavior is not as smooth. It is possible to obtain decent fits of the data of Fig. \ref{fig:lnrl} with 
the parametric form
\begin{equation}
\ln|r_q|\simeq A+B\ln (q!)+qC\ ,
\end{equation}
which is compatible with with Eq. (\ref{eq:rrate}). However, the lack of smoothness makes the discrimination 
against other behavior difficult.

\begin{figure}
\vskip60pt
\includegraphics[width=3.2in,angle=0]{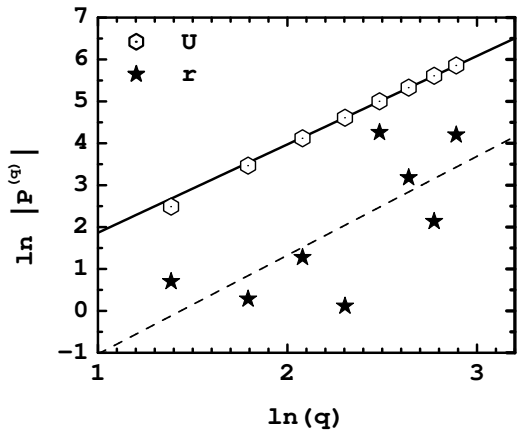}%
\vskip-60pt
\caption{Values of $\ln |P^{(2l)}|$ versus $\ln 2l$. The stars correspond to ratios of $r_{2l}$ as defined 
in Eq. (\ref{eq:p2ldef}). The hexagons correspond to the same quantity but with $r_{2l}$ replaced by
$U^{(2l)*}$.
The lines are linear fits, the slope of the solid line is 2.11 while the slope of the dashed line is 2.36.
 \label{fig:lnrat}}
\end{figure}

A better way to identify the asymptotic behavior consists in studying the ratios 
\begin{equation}
P^{(2l)}\equiv \frac{r_{2l+2}}{r_{2l}}\ .
\label{eq:p2ldef}
\end{equation}
The factorial rate of Eq. (\ref{eq:rrate}) would imply that for $l$ large enough,
\begin{equation}
P^{(2l)}\propto l^2 \ ,
\label{eq:p2l}
\end{equation}
which means a line of slope 2 in a log-log plot. 
Such a plot is provided in Fig. \ref{fig:lnrat}. 
The data is quite scattered, however, the overall rate of growth seems compatible with Eq. (\ref{eq:p2l}), the 
slope of the linear fit being 2.36. Note that without the last four data 
points, one might be tempted to conclude that the ratios are constant (this 
would imply an exponential growth rather than a factorial one).
The same quantity with $r_{2l}$ replaced by
$U^{(2l)*}$ is also provided for comparison. In this case, the linear behavior is quite evident and the slope slowly decreases from 2.22 to 2.11 as we remove
one by one  the first five data points as was done in section \ref{sec:lead}.

Our values of $r_{2l}$ are not very different from $r_6=2.048(5)$, $r_8=2.28(8)$ and $r_{10}=-13(4)$ obtained by CPRV with an improved high-temperature method or 
the values obtained by other authors with other methods   
(see  Table VIII in CPRV for details). This can be explained 
from the fact that even though Dyson's model belongs to a different 
class of universality, the critical exponents are not very different. This 
is encouraging for the possibility of improving the hierarchical approximation.
Estimates of $r_{12}$ and $r_{14}$ can be obtained from Ref. \cite{morris96}.
Using the translation $r_{2l}=(2l-1)!F_{2l-1}$, we obtain $r_{12}=38(16)$ 
with a sharp cutoff LPA and, $r_{12}=20(12)$ and $r_{14}=560(370)$ at lowest order 
in the derivative expansion. It is also interesting to compare the values 
$F_5=0.01679$, $F_7=5.317\times 10^{-4}$ and $F_9=-2.646\times 10^{-5}$ obtained with our data with the various entries of Table 7 in Ref. \cite{Morris97}. Again, the values have the same order of magnitude, but 
no precise correspondence exists with any of the approximations listed.

\section{Conclusions}
We have calculated numerically the leading and subleading amplitudes corresponding to 4  
randomly chosen local measures for the $2l$-point functions 
up to $l$ =10. We found good evidence for approximate universal relations which allow, for the 
model considered here, to 
predict the amplitudes of all the connected 
$2l$-point functions in terms of the 
amplitude for the 2-point function. We found clear indication that the universal amplitudes associated with the 
connected $2l$-point function grow as $(2l)!$.
In the context of perturbation theory, this growth seems to be related to 
the asymptotic nature of the expansion \cite{goldberg90,cornwall90,zakharov91,voloshin92}. 
However, our result is completely nonperturbative and indicates that the expansion of the generating 
function of the connected functions in power of the external field has a finite 
radius of convergence (since this generating function, as the universal function $A(z)$ of 
Eq. (\ref{eq:aofz}), has $1/(2l)!$ factors in its definition).

These results can also be used to calculate universal coefficients, denoted $r_{2l}$, appearing 
in the effective potential and which have been calculated with a variety of methods \cite{Tetradis94,Tsypin94,Guida97,Morris97,Pelissetto98b,Campostrini99,Campostrini00,Pelissetto00} for 
the universality class of the 3D Ising model with nearest neighbor interactions. 
Our results are compatible with universality with at least 5 significant digits. They are of the same order of magnitude as the existing estimates for 
other models.
To the best of our 
knowledge, we have presented the first numerical estimates of $r_{16}$, $r_{18}$ and $r_{20}$. 
They are compatible with a factorial growth of the $r_{2l}$ and the expectation 
that the 
universal function $A(z)$ has a finite radius of convergence \cite{Campostrini99}.
 
Our model does not pretend to be realistic. The fact that the non-perturbative continuum 
limit in this simplified case leads to a situation where we have high predictivity 
should be seen as an encouragement to look for nontrivial fixed point in more 
realistic theories.
The application of the method to four dimensional models requires further 
investigation.
The common wisdom is that in the pure scalar case, there is no 
non-trivial fixed point in 4D \cite{wilson74,luscher87} (see however the 
discussion of Refs. \cite{halpern95,halpern96,morris96b}).
On the other hand, for 4D models involving fermions and gauge fields, 
the question is more complex and stretches the limits of our present computational
abilities (see for instance the discussion of compact abelian fields 
coupled to fermions \cite{kogut02} or BCS inspired models of top condensation
 \cite{bardeen90}).
As the Higgs sector will soon be probed at unexplored energies, 
a special effort should be made to understand these questions.

\begin{acknowledgments}
Y. M. was supported in part by the Department of Energy
under Contract No. FG02-91ER40664 and also by a 
Faculty Scholar Award at The University of Iowa and a residential
appointment at the Obermann Center for Advanced Studies at the
University of Iowa. M. B. O. is 
supported by the Department of Energy
under Contract No. FG02-91ER40677
\end{acknowledgments}
\appendix*
\section{Numerical results for particular measures}
In this appendix, we provide numerical results obtained with the 
measures given in Table \ref{table:measures} and the Ising measure. 
\begin{table}[h]
\caption{Measures used for particular calculations. \label{table:measures}}
\renewcommand{\tabcolsep}{0.6pc} 
\renewcommand{\arraystretch}{1.6} 
\begin{tabular}{||c|c|c||} 
\hline
\hline
LG(I)  & Exp[-0.5$\phi^2$-10$\phi^4$]  & $\beta_c$=7.7036412465997630 \\
\hline
LG(II) & Exp[-2$\phi^2$-0.2$\phi^6$]  & $\beta_c$=3.1273056619243551 \\
\hline
LG(III)  & Exp[-0.1$\phi^2$-0.4$\phi^8$] & $\beta_c$=2.2259466376795976 \\
\hline
\hline
\end{tabular}
\end{table}
\begin{table*}
\caption{$a^{(q)}/a^{(2)}$ Ratios for LG measures \label{table:sub}}
\renewcommand{\tabcolsep}{0.7pc} 
\renewcommand{\arraystretch}{1.7} 
\begin{tabular}{||c||c|c||c|c||c|c||} 
\hline
\hline
    & \multicolumn{2}{c||}{LG(I)}
    & \multicolumn{2}{c||}{LG(II)} & \multicolumn{2}{c||}{LG(III)} \\
\hline
$q$ & $|a^{(q)}|$  & $a^{(q)}/a^{(2)}$ & $|a^{(q)}|$  & $a^{(q)}/a^{(2)}$ 
    & $|a^{(q)}|$    & $a^{(q)}/a^{(2)}$ \\
\hline
2   & 0.10424 & 1.00000 & 0.21404 & 1.00000 
    & 0.31178 & 1.00000 \\ 
4   & 0.21156 & 2.02961 & 0.43401 & 2.02776
    & 0.63339 & 2.03152 \\
6   & 0.33980 & 3.25988 & 0.69834 & 3.26266
    & 1.10602 & 3.25877 \\
8   & 0.47075 & 4.51615 & 0.96793 & 4.52219
    & 1.04700 & 4.51281 \\
10  & 0.60271 & 5.78214 & 1.24482 & 5.81582 
    & 1.80103 & 5.77662 \\
12  & 0.73514 & 7.05257 & 1.51927 & 7.09805
    & 2.19654 & 7.04517 \\
14  & 0.86783 & 8.32554 & 1.79472 & 8.38498 
    & 2.59268 & 8.31573 \\
16  & 1.00068 & 9.60004 & 2.07087 & 9.67519
    & 2.98897 & 9.58678 \\
18  & 1.13361 & 10.8753 & 2.34760 & 10.9681
    & 3.38638 & 10.8615\\
20  & 1.26664 & 12.1515 & 2.62528 & 12.2654 
    & 3.78335 & 12.1347 \\
\hline
\hline
\end{tabular}
\end{table*}

\begin{table*}
\caption{$a^{(q)}/a^{(2)}$ Ratios for the Ising measure\label{table:ising}}
\renewcommand{\tabcolsep}{0.7pc} 
\renewcommand{\arraystretch}{1.7} 
\begin{tabular}{||c|c|c||c|c|c||c|c|c||}
\hline
$q$ & $a^{(q)}$  & $a^{(q)}/a^{(2)}$ & $q$ & $a^{(q)}$ & $a^{(q)}/a^{(2)}$ & $q$ & $a^{(q)}$ & $a^{(q)}/a^{(2)}$ \\
\hline
2 & 0.5614  & 1.00000   & 10 & 3.24785 & 5.78527 & 18 & 6.11352 & 10.88978 \\
4 & 1.13943 & 2.02962   & 12 & 3.96229 & 7.05787 & 20 & 6.83208 & 12.16971 \\
6 & 1.83040 & 3.26042   & 14 & 4.66892 & 8.31657 &    &         &  \\
8 & 2.53632 & 4.51785   & 16 & 5.39558 & 9.61094 &    &         &  \\
\hline
\hline
\end{tabular}
\end{table*}
\begin{table*}
\caption{Dimensionless couplings $U^{(q)\star}$\label{table:ustar}}
\renewcommand{\tabcolsep}{0.8pc} 
\renewcommand{\arraystretch}{1.5} 
\begin{tabular}{||c|c|c|c|c||}
\hline
\hline
$ q $ & LG(I) & LG(II) & LG(III) & 
Ising
\\
\hline
\hline
4   & 1.5058710    & 1.5058706   & 1.5058710 
    & 1.5058709 \\ 
6   & 18.10722    & 18.10721   & 18.10722 
    & 18.10722 \\ 
8   & 579.9701& 579.9698    & 579.9702 
    & 579.9702 \\ 
10  & 35653.80   & 35653.77  & 35653.80
    & 35653.80 \\ 
12  & 3.577694E6  & 3.577690E6  & 3.577694E6 
    & 3.577694E6 \\ 
14  & 5.317628E8  & 5.317622E8  & 5.317628E8 
    & 5.317627E8 \\ 
16  & 1.097204E11 & 1.097203E11 & 1.097205E11
    & 1.097204E11 \\  
18  & 3.00025E13 & 3.00024E13  & 3.00024E13
    & 3.00025E13 \\ 
20  & 1.04998E16 & 1.04997E16  & 1.04998E16
    & 1.04997E16 \\ 
\hline
\hline
\end{tabular}
\end{table*}
\begin{table*}[h]
\caption{$r_{2l}$ values for the four measures considered, their averages and estimated errors.\label{table:r2l}}
\begin{tabular}{ccccccc}
\hline
$2l$  & LGI  &  LGII & LGIII & Ising & Average& Error\\ 
\hline
6 & 2.0149752 & 2.0149751 & 2.0149752 & 2.0149752 & 2.01497516 & 
6$\times10^{-8}$ \\
8 & 2.6795292 & 2.6795279 & 2.6795295 & 2.6795289 & 2.6795289 & 
7$\times10^{-7}$ \\
10 & -9.6011836 & -9.6011888 & -9.6011824 & -9.6011849 & 9.601185  & 
3$\times10^{-6}$ \\
12& 10.768076 & 10.768118 & 10.768070 & 10.768086 & 10.76809 & 0.00002\\
14 & 763.06239 & 763.06185 & 763.06229 & 763.06192 & 763.0621 &0.0003 \\
16 & -18380.758 & -18380.885 & -18380.933 & -18380.816 & -18380.85 & 0.08 \\
18 & 155520.69 & 155541.95 & 155500.45 & 155543.79 & 155526.72 & 20.42 \\
20 & 1.0366406E7 & 1.0378039E7 & 1.0377558E7 & 1.0373698E7 
& 1.0374$\times10^{7}$ & 5$\times10^{3}$  \\
\hline
\end{tabular}
\end{table*}

\clearpage

\begin{thebibliography}{44}
\expandafter\ifx\csname natexlab\endcsname\relax\def\natexlab#1{#1}\fi
\expandafter\ifx\csname bibnamefont\endcsname\relax
  \def\bibnamefont#1{#1}\fi
\expandafter\ifx\csname bibfnamefont\endcsname\relax
  \def\bibfnamefont#1{#1}\fi
\expandafter\ifx\csname citenamefont\endcsname\relax
  \def\citenamefont#1{#1}\fi
\expandafter\ifx\csname url\endcsname\relax
  \def\url#1{\texttt{#1}}\fi
\expandafter\ifx\csname urlprefix\endcsname\relax\def\urlprefix{URL }\fi
\providecommand{\bibinfo}[2]{#2}
\providecommand{\eprint}[2][]{\url{#2}}

\bibitem[{\citenamefont{Hagiwara et~al.}(2002)}]{pdg02}
\bibinfo{author}{\bibfnamefont{K.}~\bibnamefont{Hagiwara}} \bibnamefont{et~al.}
  (\bibinfo{collaboration}{Particle Data Group}), \bibinfo{journal}{Phys. Rev.}
  \textbf{\bibinfo{volume}{D66}}, \bibinfo{pages}{010001}
  (\bibinfo{year}{2002}).

\bibitem[{\citenamefont{K.Wilson}(1972)}]{wilson72}
\bibinfo{author}{\bibnamefont{K.Wilson}}, \bibinfo{journal}{Phys.\ Rev.\ D}
  \textbf{\bibinfo{volume}{6}}, \bibinfo{pages}{419} (\bibinfo{year}{1972}).

\bibitem[{\citenamefont{Godina et~al.}(1998{\natexlab{a}})\citenamefont{Godina,
  Meurice, and Oktay}}]{gam3rapid}
\bibinfo{author}{\bibfnamefont{J.}~\bibnamefont{Godina}},
  \bibinfo{author}{\bibfnamefont{Y.}~\bibnamefont{Meurice}}, \bibnamefont{and}
  \bibinfo{author}{\bibfnamefont{M.}~\bibnamefont{Oktay}},
  \bibinfo{journal}{Phys. Rev. D} \textbf{\bibinfo{volume}{57}},
  \bibinfo{pages}{R6581} (\bibinfo{year}{1998}{\natexlab{a}}).

\bibitem[{\citenamefont{Privman et~al.}(1991)\citenamefont{Privman, Hohenberg,
  and Aharony}}]{privman91}
\bibinfo{author}{\bibfnamefont{V.}~\bibnamefont{Privman}},
  \bibinfo{author}{\bibfnamefont{P.~C.} \bibnamefont{Hohenberg}},
  \bibnamefont{and} \bibinfo{author}{\bibfnamefont{A.}~\bibnamefont{Aharony}},
  in \emph{\bibinfo{booktitle}{Phase Transitions and Critical Phenonema, Vol.
  14}}, edited by \bibinfo{editor}{\bibfnamefont{L.}~\bibnamefont{Domb}}
  \bibnamefont{and} \bibinfo{editor}{\bibfnamefont{J.}~\bibnamefont{Lebowitz}}
  (\bibinfo{publisher}{Academic Press}, \bibinfo{address}{New York},
  \bibinfo{year}{1991}), pp. \bibinfo{pages}{4--121}.

\bibitem[{\citenamefont{Campostrini et~al.}(1999)\citenamefont{Campostrini,
  Pelissetto, Rossi, and Vicari}}]{Campostrini99}
\bibinfo{author}{\bibfnamefont{M.}~\bibnamefont{Campostrini}},
  \bibinfo{author}{\bibfnamefont{A.}~\bibnamefont{Pelissetto}},
  \bibinfo{author}{\bibfnamefont{P.}~\bibnamefont{Rossi}}, \bibnamefont{and}
  \bibinfo{author}{\bibfnamefont{E.}~\bibnamefont{Vicari}},
  \bibinfo{journal}{Phys. Rev.} \textbf{\bibinfo{volume}{E60}},
  \bibinfo{pages}{3526} (\bibinfo{year}{1999}), \eprint{cond-mat/9905078}.

\bibitem[{\citenamefont{Tetradis and Wetterich}(1994)}]{Tetradis94}
\bibinfo{author}{\bibfnamefont{N.}~\bibnamefont{Tetradis}} \bibnamefont{and}
  \bibinfo{author}{\bibfnamefont{C.}~\bibnamefont{Wetterich}},
  \bibinfo{journal}{Nucl. Phys.} \textbf{\bibinfo{volume}{B422}},
  \bibinfo{pages}{541} (\bibinfo{year}{1994}), \eprint{hep-ph/9308214}.

\bibitem[{\citenamefont{Guida and Zinn-Justin}(1997)}]{Guida97}
\bibinfo{author}{\bibfnamefont{R.}~\bibnamefont{Guida}} \bibnamefont{and}
  \bibinfo{author}{\bibfnamefont{J.}~\bibnamefont{Zinn-Justin}},
  \bibinfo{journal}{Nucl. Phys.} \textbf{\bibinfo{volume}{B489}},
  \bibinfo{pages}{626} (\bibinfo{year}{1997}), \eprint{hep-th/9610223}.

\bibitem[{\citenamefont{Morris}(1997)}]{Morris97}
\bibinfo{author}{\bibfnamefont{T.~R.} \bibnamefont{Morris}},
  \bibinfo{journal}{Nucl. Phys.} \textbf{\bibinfo{volume}{B495}},
  \bibinfo{pages}{477} (\bibinfo{year}{1997}), \eprint{hep-th/9612117}.

\bibitem[{\citenamefont{Pelissetto and Vicari}(1998)}]{Pelissetto98b}
\bibinfo{author}{\bibfnamefont{A.}~\bibnamefont{Pelissetto}} \bibnamefont{and}
  \bibinfo{author}{\bibfnamefont{E.}~\bibnamefont{Vicari}},
  \bibinfo{journal}{Nucl. Phys.} \textbf{\bibinfo{volume}{B522}},
  \bibinfo{pages}{605} (\bibinfo{year}{1998}), \eprint{cond-mat/9801098}.

\bibitem[{\citenamefont{Campostrini et~al.}(2000)\citenamefont{Campostrini,
  Pelissetto, Rossi, and Vicari}}]{Campostrini00}
\bibinfo{author}{\bibfnamefont{M.}~\bibnamefont{Campostrini}},
  \bibinfo{author}{\bibfnamefont{A.}~\bibnamefont{Pelissetto}},
  \bibinfo{author}{\bibfnamefont{P.}~\bibnamefont{Rossi}}, \bibnamefont{and}
  \bibinfo{author}{\bibfnamefont{E.}~\bibnamefont{Vicari}},
  \bibinfo{journal}{Phys. Rev.} \textbf{\bibinfo{volume}{B62}},
  \bibinfo{pages}{5843} (\bibinfo{year}{2000}), \eprint{cond-mat/0001440}.

\bibitem[{\citenamefont{Tsypin}(1994)}]{Tsypin94}
\bibinfo{author}{\bibfnamefont{M.~M.} \bibnamefont{Tsypin}},
  \bibinfo{journal}{Phys. Rev. Lett.} \textbf{\bibinfo{volume}{73}},
  \bibinfo{pages}{2015} (\bibinfo{year}{1994}).

\bibitem[{\citenamefont{Pelissetto and Vicari}(2002)}]{Pelissetto00}
\bibinfo{author}{\bibfnamefont{A.}~\bibnamefont{Pelissetto}} \bibnamefont{and}
  \bibinfo{author}{\bibfnamefont{E.}~\bibnamefont{Vicari}},
  \bibinfo{journal}{Phys. Rept.} \textbf{\bibinfo{volume}{368}},
  \bibinfo{pages}{549} (\bibinfo{year}{2002}), \eprint{cond-mat/0012164}.

\bibitem[{\citenamefont{Dyson}(1969)}]{dyson69}
\bibinfo{author}{\bibfnamefont{F.}~\bibnamefont{Dyson}},
  \bibinfo{journal}{Comm.\ Math.\ Phys.} \textbf{\bibinfo{volume}{12}},
  \bibinfo{pages}{91} (\bibinfo{year}{1969}).

\bibitem[{\citenamefont{Baker}(1972)}]{baker72}
\bibinfo{author}{\bibfnamefont{G.}~\bibnamefont{Baker}},
  \bibinfo{journal}{Phys.\ Rev.\ B} \textbf{\bibinfo{volume}{5}},
  \bibinfo{pages}{2622} (\bibinfo{year}{1972}).

\bibitem[{\citenamefont{Meurice}(2003)}]{smalld03}
\bibinfo{author}{\bibfnamefont{Y.}~\bibnamefont{Meurice}},
  \bibinfo{journal}{Phys. Rev. E (in press)}  (\bibinfo{year}{2003}),
  \eprint{cond-mat/0312188}.

\bibitem[{\citenamefont{Parisi}(1988)}]{parisi88}
\bibinfo{author}{\bibfnamefont{G.}~\bibnamefont{Parisi}},
  \emph{\bibinfo{title}{Statistical Field Theory}} (\bibinfo{publisher}{Addison
  Wesley}, \bibinfo{address}{New York}, \bibinfo{year}{1988}).

\bibitem[{\citenamefont{Goldberg}(1990)}]{goldberg90}
\bibinfo{author}{\bibfnamefont{H.}~\bibnamefont{Goldberg}},
  \bibinfo{journal}{Phys. Lett.} \textbf{\bibinfo{volume}{B246}},
  \bibinfo{pages}{445} (\bibinfo{year}{1990}).

\bibitem[{\citenamefont{Cornwall}(1990)}]{cornwall90}
\bibinfo{author}{\bibfnamefont{J.~M.} \bibnamefont{Cornwall}},
  \bibinfo{journal}{Phys. Lett.} \textbf{\bibinfo{volume}{B243}},
  \bibinfo{pages}{271} (\bibinfo{year}{1990}).

\bibitem[{\citenamefont{Zakharov}(1991)}]{zakharov91}
\bibinfo{author}{\bibfnamefont{V.~I.} \bibnamefont{Zakharov}},
  \bibinfo{journal}{Phys. Rev. Lett.} \textbf{\bibinfo{volume}{67}},
  \bibinfo{pages}{3650} (\bibinfo{year}{1991}).

\bibitem[{\citenamefont{Voloshin}(1992)}]{voloshin92}
\bibinfo{author}{\bibfnamefont{M.~B.} \bibnamefont{Voloshin}},
  \bibinfo{journal}{Nucl. Phys.} \textbf{\bibinfo{volume}{B383}},
  \bibinfo{pages}{233} (\bibinfo{year}{1992}).

\bibitem[{\citenamefont{Aharony and Ahlers}(1980)}]{aharony80}
\bibinfo{author}{\bibfnamefont{A.}~\bibnamefont{Aharony}} \bibnamefont{and}
  \bibinfo{author}{\bibfnamefont{G.}~\bibnamefont{Ahlers}},
  \bibinfo{journal}{Phys. Rev. Lett.} \textbf{\bibinfo{volume}{44}},
  \bibinfo{pages}{782} (\bibinfo{year}{1980}).

\bibitem[{\citenamefont{Chang and Houghton}(1980)}]{chang80}
\bibinfo{author}{\bibfnamefont{M.}~\bibnamefont{Chang}} \bibnamefont{and}
  \bibinfo{author}{\bibfnamefont{A.}~\bibnamefont{Houghton}},
  \bibinfo{journal}{Phys. Rev. Lett.} \textbf{\bibinfo{volume}{44}},
  \bibinfo{pages}{785} (\bibinfo{year}{1980}).

\bibitem[{\citenamefont{Wegner}(1976)}]{wegner76}
\bibinfo{author}{\bibfnamefont{F.~J.} \bibnamefont{Wegner}}, in
  \emph{\bibinfo{booktitle}{Phase Transitions and Critical Phenonema, Vol. 6}},
  edited by \bibinfo{editor}{\bibfnamefont{L.}~\bibnamefont{Domb}}
  \bibnamefont{and} \bibinfo{editor}{\bibfnamefont{M.~S.} \bibnamefont{Green}}
  (\bibinfo{publisher}{Academic Press}, \bibinfo{address}{New York},
  \bibinfo{year}{1976}), pp. \bibinfo{pages}{7--124}.

\bibitem[{\citenamefont{Koch and Wittwer}(1995)}]{koch95}
\bibinfo{author}{\bibfnamefont{H.}~\bibnamefont{Koch}} \bibnamefont{and}
  \bibinfo{author}{\bibfnamefont{P.}~\bibnamefont{Wittwer}},
  \bibinfo{journal}{Math. Phys. Electr. Jour.} \textbf{\bibinfo{volume}{1}},
  \bibinfo{pages}{Paper 6} (\bibinfo{year}{1995}).

\bibitem[{\citenamefont{Godina et~al.}(1999)\citenamefont{Godina, Meurice, and
  Oktay}}]{gam3}
\bibinfo{author}{\bibfnamefont{J.}~\bibnamefont{Godina}},
  \bibinfo{author}{\bibfnamefont{Y.}~\bibnamefont{Meurice}}, \bibnamefont{and}
  \bibinfo{author}{\bibfnamefont{M.}~\bibnamefont{Oktay}},
  \bibinfo{journal}{Phys. Rev. D} \textbf{\bibinfo{volume}{59}},
  \bibinfo{pages}{096002} (\bibinfo{year}{1999}).

\bibitem[{\citenamefont{Godina et~al.}(1998{\natexlab{b}})\citenamefont{Godina,
  Meurice, Oktay, and Niermann}}]{finite}
\bibinfo{author}{\bibfnamefont{J.}~\bibnamefont{Godina}},
  \bibinfo{author}{\bibfnamefont{Y.}~\bibnamefont{Meurice}},
  \bibinfo{author}{\bibfnamefont{M.}~\bibnamefont{Oktay}}, \bibnamefont{and}
  \bibinfo{author}{\bibfnamefont{S.}~\bibnamefont{Niermann}},
  \bibinfo{journal}{Phys. Rev. D} \textbf{\bibinfo{volume}{57}},
  \bibinfo{pages}{6326} (\bibinfo{year}{1998}{\natexlab{b}}).

\bibitem[{\citenamefont{Wilson}(1971)}]{wilson71b}
\bibinfo{author}{\bibfnamefont{K.}~\bibnamefont{Wilson}},
  \bibinfo{journal}{Phys. Rev. B.} \textbf{\bibinfo{volume}{4}},
  \bibinfo{pages}{3185} (\bibinfo{year}{1971}).

\bibitem[{\citenamefont{Meurice and Ordaz}(1996)}]{fam}
\bibinfo{author}{\bibfnamefont{Y.}~\bibnamefont{Meurice}} \bibnamefont{and}
  \bibinfo{author}{\bibfnamefont{G.}~\bibnamefont{Ordaz}}, \bibinfo{journal}{J.
  Phys. A (Letter to the Editor)} \textbf{\bibinfo{volume}{29}},
  \bibinfo{pages}{L635} (\bibinfo{year}{1996}).

\bibitem[{\citenamefont{Godina et~al.}(2000)\citenamefont{Godina, Meurice, and
  Oktay}}]{hyper}
\bibinfo{author}{\bibfnamefont{J.~J.} \bibnamefont{Godina}},
  \bibinfo{author}{\bibfnamefont{Y.}~\bibnamefont{Meurice}}, \bibnamefont{and}
  \bibinfo{author}{\bibfnamefont{M.}~\bibnamefont{Oktay}},
  \bibinfo{journal}{Phys. Rev. D} \textbf{\bibinfo{volume}{61}},
  \bibinfo{pages}{114509} (\bibinfo{year}{2000}).

\bibitem[{\citenamefont{Niemeijer and van Leeuwen}(1976)}]{niemeijer76}
\bibinfo{author}{\bibfnamefont{T.}~\bibnamefont{Niemeijer}} \bibnamefont{and}
  \bibinfo{author}{\bibfnamefont{J.}~\bibnamefont{van Leeuwen}}, in
  \emph{\bibinfo{booktitle}{Phase Transitions and Critical Phenomena, vol. 6}},
  edited by \bibinfo{editor}{\bibfnamefont{C.}~\bibnamefont{Domb}}
  \bibnamefont{and} \bibinfo{editor}{\bibfnamefont{M.}~\bibnamefont{Green}}
  (\bibinfo{publisher}{Academic Press}, \bibinfo{address}{New York},
  \bibinfo{year}{1976}).

\bibitem[{\citenamefont{Meurice et~al.}(1995)\citenamefont{Meurice, Ordaz, and
  Rodgers}}]{osc1}
\bibinfo{author}{\bibfnamefont{Y.}~\bibnamefont{Meurice}},
  \bibinfo{author}{\bibfnamefont{G.}~\bibnamefont{Ordaz}}, \bibnamefont{and}
  \bibinfo{author}{\bibfnamefont{V.~G.~J.} \bibnamefont{Rodgers}},
  \bibinfo{journal}{Phys.\ Rev.\ Lett.} \textbf{\bibinfo{volume}{75}},
  \bibinfo{pages}{4555} (\bibinfo{year}{1995}).

\bibitem[{\citenamefont{Meurice et~al.}(1997)\citenamefont{Meurice, Niermann,
  and Ordaz}}]{osc2}
\bibinfo{author}{\bibfnamefont{Y.}~\bibnamefont{Meurice}},
  \bibinfo{author}{\bibfnamefont{S.}~\bibnamefont{Niermann}}, \bibnamefont{and}
  \bibinfo{author}{\bibfnamefont{G.}~\bibnamefont{Ordaz}},
  \bibinfo{journal}{J.\ Stat.\ Phys.} \textbf{\bibinfo{volume}{87}},
  \bibinfo{pages}{363} (\bibinfo{year}{1997}).

\bibitem[{\citenamefont{Oktay}()}]{oktayphd}
\bibinfo{author}{\bibfnamefont{M.~B.} \bibnamefont{Oktay}},
  \emph{\bibinfo{title}{Nonperturbative methods for hierarchical models}},
  \bibinfo{note}{(Ph. D. thesis), UMI-30-18602}.

\bibitem[{\citenamefont{Glazek and Wilson}(2002)}]{wilson03}
\bibinfo{author}{\bibfnamefont{S.~D.} \bibnamefont{Glazek}} \bibnamefont{and}
  \bibinfo{author}{\bibfnamefont{K.~G.} \bibnamefont{Wilson}},
  \bibinfo{journal}{Phys. Rev. Lett.} \textbf{\bibinfo{volume}{89}},
  \bibinfo{pages}{230401} (\bibinfo{year}{2002}), \eprint{hep-th/0203088}.

\bibitem[{\citenamefont{Braaten and Hammer}(2003)}]{braaten03}
\bibinfo{author}{\bibfnamefont{E.}~\bibnamefont{Braaten}} \bibnamefont{and}
  \bibinfo{author}{\bibfnamefont{H.~W.} \bibnamefont{Hammer}},
  \bibinfo{journal}{Phys. Rev. Lett.} \textbf{\bibinfo{volume}{91}},
  \bibinfo{pages}{102002} (\bibinfo{year}{2003}), \eprint{nucl-th/0303038}.

\bibitem[{\citenamefont{Glimm and Jaffe}(1987)}]{glimm87}
\bibinfo{author}{\bibfnamefont{J.}~\bibnamefont{Glimm}} \bibnamefont{and}
  \bibinfo{author}{\bibfnamefont{A.}~\bibnamefont{Jaffe}},
  \emph{\bibinfo{title}{Quantum Physics}}
  (\bibinfo{publisher}{Springer-Verlag}, \bibinfo{address}{New York},
  \bibinfo{year}{1987}).

\bibitem[{\citenamefont{Morris}(1996{\natexlab{a}})}]{morris96}
\bibinfo{author}{\bibfnamefont{T.~R.} \bibnamefont{Morris}},
  \bibinfo{journal}{Nucl. Phys.} \textbf{\bibinfo{volume}{B458}},
  \bibinfo{pages}{477} (\bibinfo{year}{1996}{\natexlab{a}}),
  \eprint{hep-th/9508017}.

\bibitem[{\citenamefont{Wilson and Kogut}(1974)}]{wilson74}
\bibinfo{author}{\bibfnamefont{K.}~\bibnamefont{Wilson}} \bibnamefont{and}
  \bibinfo{author}{\bibfnamefont{J.}~\bibnamefont{Kogut}},
  \bibinfo{journal}{Phys.\ Rep.} \textbf{\bibinfo{volume}{12}},
  \bibinfo{pages}{75} (\bibinfo{year}{1974}).

\bibitem[{\citenamefont{Luscher and Weisz}(1987)}]{luscher87}
\bibinfo{author}{\bibfnamefont{M.}~\bibnamefont{Luscher}} \bibnamefont{and}
  \bibinfo{author}{\bibfnamefont{P.}~\bibnamefont{Weisz}},
  \bibinfo{journal}{Nucl. Phys.} \textbf{\bibinfo{volume}{B290}},
  \bibinfo{pages}{25} (\bibinfo{year}{1987}).

\bibitem[{\citenamefont{Halpern and Huang}(1995)}]{halpern95}
\bibinfo{author}{\bibfnamefont{K.}~\bibnamefont{Halpern}} \bibnamefont{and}
  \bibinfo{author}{\bibfnamefont{K.}~\bibnamefont{Huang}},
  \bibinfo{journal}{Phys. Rev. Lett.} \textbf{\bibinfo{volume}{74}},
  \bibinfo{pages}{3526} (\bibinfo{year}{1995}), \eprint{hep-th/9406199}.

\bibitem[{\citenamefont{Halpern and Huang}(1996)}]{halpern96}
\bibinfo{author}{\bibfnamefont{K.}~\bibnamefont{Halpern}} \bibnamefont{and}
  \bibinfo{author}{\bibfnamefont{K.}~\bibnamefont{Huang}},
  \bibinfo{journal}{Phys. Rev. Lett.} \textbf{\bibinfo{volume}{77}},
  \bibinfo{pages}{1659} (\bibinfo{year}{1996}).

\bibitem[{\citenamefont{Morris}(1996{\natexlab{b}})}]{morris96b}
\bibinfo{author}{\bibfnamefont{T.~R.} \bibnamefont{Morris}},
  \bibinfo{journal}{Phys. Rev. Lett.} \textbf{\bibinfo{volume}{77}},
  \bibinfo{pages}{1658} (\bibinfo{year}{1996}{\natexlab{b}}),
  \eprint{hep-th/9601128}.

\bibitem[{\citenamefont{Kogut and Strouthos}(2003)}]{kogut02}
\bibinfo{author}{\bibfnamefont{J.~B.} \bibnamefont{Kogut}} \bibnamefont{and}
  \bibinfo{author}{\bibfnamefont{C.~G.} \bibnamefont{Strouthos}},
  \bibinfo{journal}{Phys. Rev.} \textbf{\bibinfo{volume}{D67}},
  \bibinfo{pages}{034504} (\bibinfo{year}{2003}), \eprint{hep-lat/0211024}.

\bibitem[{\citenamefont{Bardeen et~al.}(1990)\citenamefont{Bardeen, Hill, and
  Lindner}}]{bardeen90}
\bibinfo{author}{\bibfnamefont{W.~A.} \bibnamefont{Bardeen}},
  \bibinfo{author}{\bibfnamefont{C.~T.} \bibnamefont{Hill}}, \bibnamefont{and}
  \bibinfo{author}{\bibfnamefont{M.}~\bibnamefont{Lindner}},
  \bibinfo{journal}{Phys. Rev.} \textbf{\bibinfo{volume}{D41}},
  \bibinfo{pages}{1647} (\bibinfo{year}{1990}).

\end{thebibliography}

\end{document}